\documentclass[%
reprint,
 amsmath,amssymb,
pra,%draft,
]{revtex4-1}

\usepackage{graphicx}% Include figure files

\usepackage{siunitx}% Include figure files
\usepackage{dcolumn}% Align table columns on decimal point
\usepackage{bm}% bold math
\usepackage[hidelinks]{hyperref}% add hypertext capabilities
\usepackage[mathlines]{lineno}% Enable numbering of text and display math
\begin{document}

\preprint{}

\title{Autonomous deployment of a solar panel using an Elastic Origami and Distributed Shape Memory Polymer Actuators}%

\author{Tian Chen$^a$, Osama R. Bilal$^b$, Robert Lang$^c$, Chiara Daraio$^{b*}$, Kristina Shea$^{a}$}
\email{To whom correspondence may be addressed. Email: daraio@caltech.edu or kshea@ethz.ch}
\affiliation{$^a$ Department of Mechanical and Process Engineering, ETH Zurich, Zurich 8092, Switzerland}
\affiliation{$^b$ Division of Engineering and Applied Science, California Institute of Technology, Pasadena, CA 91125, USA}
\affiliation{$^c$ Lang Origami, Alamo, CA 94507, USA}
\date{\today}

\begin{abstract}
Deployable mechanical systems such as space solar panels rely on the intricate stowage of passive modules, and sophisticated deployment using a network of motorized actuators. As a result, a significant portion of the stowed mass and volume are occupied by these support systems. An autonomous solar panel array deployed using the inherent material behavior remains elusive. In this work, we develop an autonomous self-deploying solar panel array that is programmed to activate in response to changes in the surrounding temperature. We study an elastic ``flasher'' origami sheet embedded in a circle of scissor mechanisms, both printed with shape memory polymers. The scissor mechanisms are optimized to provide the maximum expansion ratio while delivering the necessary force for deployment. The origami sheet is also optimized to carry the maximum number of solar panels given space constraints. We show how the folding of the ``flasher'' origami exhibits a bifurcation behavior resulting in either a cone or disk shape both numerically and in experiments. A folding strategy is devised to avoid the undesired cone shape. The resulting design is entirely 3D printed, achieves an expansion ratio of 1000\% in under 40 seconds, and shows excellent agreement with simulation prediction both in the stowed and deployed configurations.
\end{abstract}

\maketitle

\section{Introduction}
The surging demand for deployable mechanical systems is predominantly driven by the need to explore ever more inaccessible environments and to deliver increasingly large and complex payloads. Existing need for deployable systems includes space-based solar power~\cite{McSpadden2002}~\cite{McGuire2016}, geoengineering~\cite{Sigel2014}, antennas~\cite{Zhang2014}, and propulsion~\cite{Mori2010}. The design of these mechatronic systems largely considers the stowage of the payload~\cite{Schenk2013} and the actuation which transform the payload from the stowed to the deployed configuration~\cite{Block2011}. Traditionally, the actuation is achieved through a network of controlled actuators, which unpacks the passive payload. While numerous studies are done to ensure a robust stowage and deployment process, with ever more complex systems, autonomous large-scale deployment, weight and volume of power supply, component jamming and failure, placement of the actuators, and advanced control mechanisms continue to challenge engineers.

The nascent field of programmable matter, where engineers integrate material and functionality~\cite{Yang2017} to create more reliable, adaptive and robust designs, is utilized here to address these challenges. Programmable matter refers to a material or a structure whose shape or stiffness can be controlled~\cite{Hawkes2010, bilal2017reprogrammable}. Numerous designs have started to demonstrate shape reconfiguration~\cite{Chen2016c}. For example, fabricating swelling hydrogel to create doubly curved surfaces~\cite{Raviv2014} or flower petals~\cite{Gladman2016}, programming and activation of shape memory polymer to transform a sheet into a box~\cite{Mao2015} or to deploy a shape that was stowed in a cylinder~\cite{Wagner2017} and transformable sheets using liquid crystal polymer~\cite{White2015} or printing pre-strain~\cite{Bauhofer2017}. When the objects are fully 3D printed, the area is also called 4D printing, i.e. objects that reconfigure themselves in time.

 % A negative Poisson's ratio meta material has been used for vibration mitigation~\cite{Chen2017d}. 
 
Functionalities beyond shape reconfiguration have also been demonstrated. Pneumatics have been used to actuate soft crawlers~\cite{Shepherd2011}, a robotic octopus~\cite{Wehner2016}, and transformable surfaces~\cite{Pikul2017}.Electrical components have been embedded within the material itself as well to create soft electronics~\cite{Valentine2017} and cardiac micro-physiological devices~\cite{Lind2016}. Instability has been used to create multi-stable structures without mechanical hinges~\cite{Chen2017a}, the actuation of soft directional swimmers~\cite{Chen2017b}, and stable propagation of elastic energy~\cite{Raney2016}. While these works eliminate the need for a physical tether or rigid machineries, the active components are still being treated as separate entities that need to be embedded into a passive vessel. This complicates the design process and makes the system error prone, e.g. jamming.

\begin{figure}[!ht]
	\includegraphics[width=0.45\textwidth]{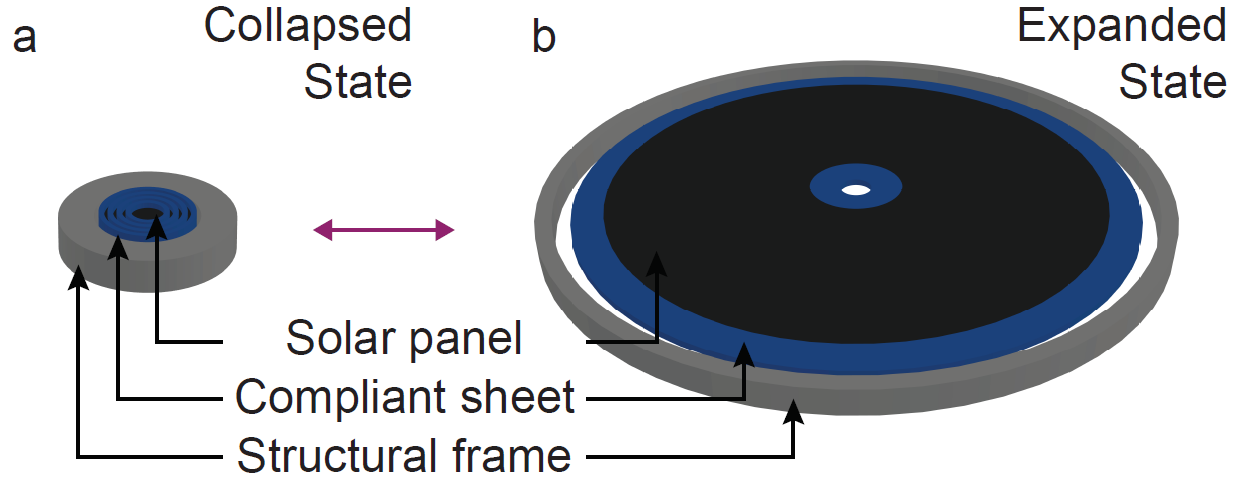}
	\caption{The schematic of the deployment strategy from (a) to (b) of the presented mechanical system. An outer ring acts as the primary actuation mechanism and structural support. An inner substrate forms the secondary actuation mechanism and provides the surface to carry the solar panels. Both components are fabricated using a shape memory polymer as the actuation mechanism.}
	\label{fig:1}
\end{figure}

By utilizing advanced material systems, it becomes possible to embed and distribute sensing, control, actuation, and logic within the material itself rather than as discrete components in an assembly~\cite{Truby2016}. Using a self-deploying soft solar panel array as an example, we aim to demonstrate stable self-deployment through distributed actuation using the shape memory effect. The proposed mechanical system consists of an outer ring and an inner substrate (\ref{fig:1}a,b). The outer ring serves as the primary structural system and actuation mechanism. The inner substrate carries the solar panels and serves as the secondary actuation mechanism. In such a configuration, we show that the inner substrate and the solar panels fit entirely within the inevitable void of the collapsed outer ring.

To achieve a tunable expansion ratio, the outer ring adopts the design of the Hoberman Sphere~\cite{Hoberman1990} where a series of scissor mechanisms in a ring configuration serves as the structural system and primary actuator. An elastic ``flasher'' origami is used as a deployable surface and a secondary actuator. We pose the maximization of expansion ratio and the maximization of the number of fitted solar panels as discrete problems that are solved analytically considering only kinematics. The resulting mechanical system is able to achieve an area expansion ratio of ten, from approximately \num{0.05} to \SI{0.5}{\metre\squared}. The dynamics of folding the ``flasher'' origami is studied in detail as a bifurcation behavior is observed in both simulation and experiments. It is shown that depending on the rate of folding, the origami folds into either a cone or a disk shape. As a result, a rotational mechanism is devised for collapsing of the whole system into its stowed configuration.

For actuation, by distributing a shape memory polymer in both the Hoberman ring and in the origami substrate, we enable tuning of material stiffness by sensing the surrounding temperature. In both stowage and operating configurations where the temperature is lower than the glass transition temperature ($T_{\mathrm{g}}$) of the polymer, both components contribute to the system's overall stiffness. During collapsing and deployment, sensing a temperature higher than ($T_{\mathrm{g}}$), the material reduces in stiffness by orders of magnitude and becomes compliant. In this example, collapsing to the stowed configuration is manual and programs a pre-strain in the shape memory polymers. The autonomous self-deployment occurs when the reduction in stiffness relaxes the pre-strain and reconfigures the system.

\begin{figure}[!ht]
	\includegraphics[width=0.45\textwidth]{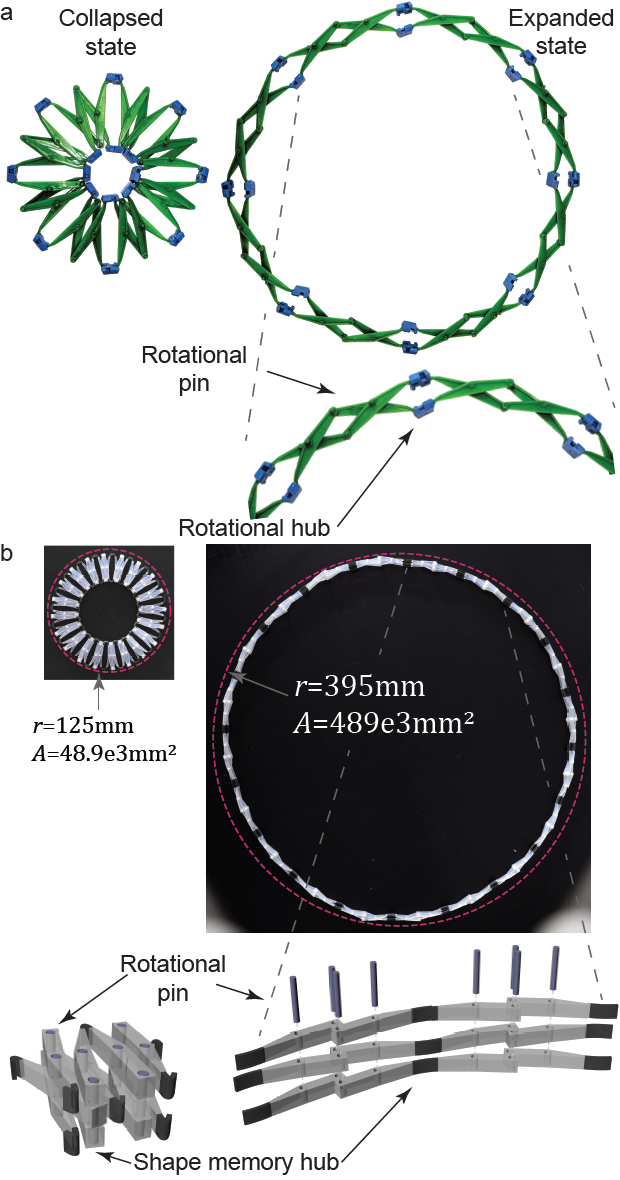}
	\caption{The design of the autonomous self-deploying Hoberman ring. a) A ring of the Hoberman Sphere toy in both collapsed and expanded configuration. b) Zoomed to show the mechanism with hubs and pins indicated. c) Fabricated design of the 3D printed Hoberman ring in both collapsed and expanded configurations showing 10 times change in area. d) Zoomed in render showing the shape memory polymer hubs and articulated pins.}
	\label{fig:2}
\end{figure}

\section{Results and discussion}
The design and functionality of the self-deploying solar panel array is described in detail first through its components, i.e., the Hoberman ring, and the elastic ``flasher'' origami. Then, the behavior of the entire assembly is presented. The Hoberman ring is described in terms of its 1) mechanism, 2) analysis of its ratio of expansion, and 3) means of actuation. The elastic ``flasher'' origami is then described with respect to 1) the parametric optimization of the crease pattern, 2) dynamics of folding that result in a bifurcation behavior, 3) means of actuation. The description of the entire assembly consists of the self-deploying experiment which shows two distinct phases of deployment.

\subsection{Hoberman Ring}
% Intro
The Hoberman Sphere~\cite{Hoberman1990} was initially patented by Hoberman before being popularized as a toy. It makes clever use of the classic scissor mechanism~\cite{Larson1966} to create a sphere-like object which can expand to several times its collapsed volume (Fig.~\ref{fig:2}a). Hubs are placed at the intersection of the scissors, and pins are used to enable the scissoring mechanism (Fig.~\ref{fig:2}b).

\begin{figure}[!ht]
	\includegraphics[width=0.45\textwidth]{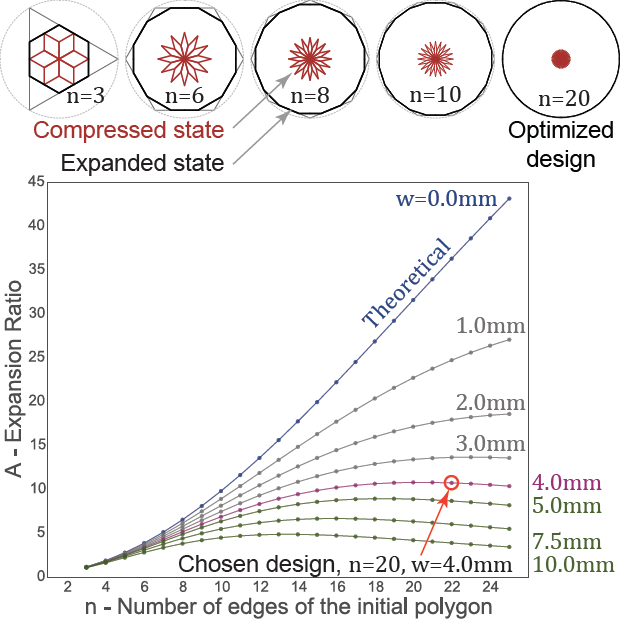}
	\caption{Parametric design of the Hoberman ring accounting for fabrication limitations and width of the members. Expansion ratio as a function of number of edges of the initial polygon plotted for different member thickness, $w$. For member thickness $w=4$, the optimal expansion ratio occurs at $n=20$. Derivation can be found in the Appendix.}
	\label{fig:3}
\end{figure}

% Kinematics
We adopt a 2D version (i.e., ring) of this design to make an expanding planar mechanism (Fig.~\ref{fig:2}c). Inspired by numerous works which convert pure mechanisms into mechanical meta-materials~\cite{Bertoldi2017, Janbaz2018}, we enable autonomous expansion within the ring by replacing the hubs with shape memory polymers as actuators (Fig.~\ref{fig:2}d). The shape memory hubs have two polymer states, glassy or rubbery, depending on whether they are above or below the glass transition temperature, $T_\mathrm{g}$. At a temperature above $T_\mathrm{g}$, the hubs can be deformed in a predefined shape. If the hubs are held in this shape while cooled down, they retrain the shape conferred upon heating, by ``locking in'' the strains. When reheated past $T_\mathrm{g}$, the shape memory hubs relax this pre-strain and consequently deploy the structure.

The Hoberman ring is constructed using a regular polygon as the base geometry. The number of edges, $n$, of this polygon determines the expansion ratio of the mechanism given a set of physical and fabrication constraints. We define the expansion ratio as $\Delta A=A_\mathrm{expanded}/A_\mathrm{collapsed}$. The radius of each configuration is measured as the distance between the polygon center and the outer most vertex of the geometry. To maximize this expansion ratio, we derive the relationship between $\Delta A$ against $n$. The schematics (Fig.~\ref{fig:3} top) show the geometric configuration of collapsed and expanded rings, for different number of edges, $n$, assuming zero thickness, $w$, of the members. In this theoretical scenario, the inner vertices of the ring coincides at the center of the polygon~\cite{You1997}~\cite{Patel2007}. By introducing finite thickness, the inner vertices collide with each other as the ring collapses, forming an inner circular void. Quantitatively, it is observed that for a given $w>0$ with increasing $n$, the expansion ratio reaches a maximum (Fig.~\ref{fig:3} bottom). This maximum occurs with smaller $n$ for larger values of $w$ as the thickness of the members prevents the mechanism from collapsing to the center. Accounting for fabrication limitations and robustness to repeated cycles of collapse-expansion, we choose a thickness of $w=\SI{4}{\milli\metre}$ for testing. For this thickness, the maximum expansion ratio of $\Delta A=10.79$ occurs for $n=20$ edges.

\begin{figure}[!ht]
	\includegraphics[width=0.45\textwidth]{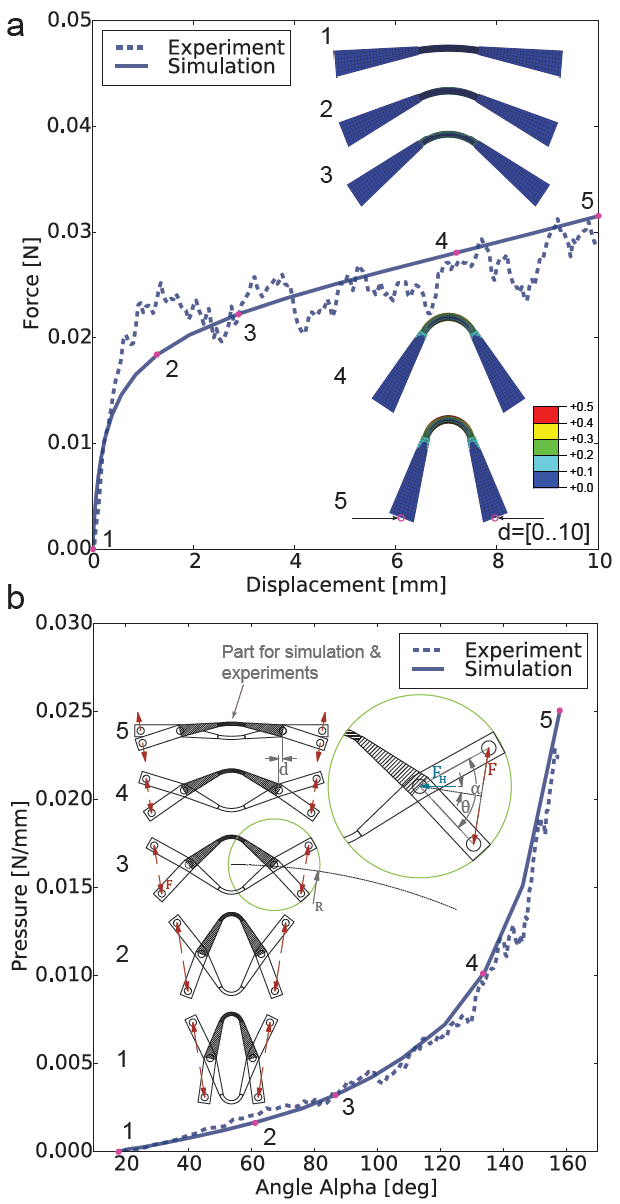}
	\caption{Mechanics of the shape memory hubs and the relationship with the radial pressure of the Hoberman ring. a)Force displacement plot of a single shape memory hub from experiments and simulations. b) Overall pressure produced by all the shape memory joints of a $n=20$ Hoberman ring as a function of the angle in the scissor mechanism.
}
	\label{fig:4}
\end{figure}

The optimized ring including the SMPs is fabricated in the expanded configuration and programmed to the collapsed configuration. As the Hoberman mechanism has a Poisson's Ratio of $-1$ at all strains, the programming is done under $T\ge T_{\mathrm{g}}$ by uniformly compressing the ring until it fits into a predefined mold. The mold is dimensioned such that the area expansion ratio is 10, which is slightly less than the predicted value (10.79). The specimen is cooled while in the mold and then extracted. The fabricated specimens demonstrate the predicted area change as well as structural stability in both collapsed and expanded configurations.

The radial pressure required during compression and exerted during expansion can be predicted by understanding the force displacement behavior of the shape memory hubs. It is known that the recovery force is dependent on the dimensions of the shape memory material, strain rate, and the ambient temperature~\cite{Wagner2017}. Compression experiments are done to a segment of the Hoberman ring under temperature $T>T_\mathrm{g}$ with \SI{10}{\milli\metre} of displacement being prescribed to both sides of the segment at a rate of \SI{0.5}{\milli\metre\per\second}. The ends of the segment are pin-connected to the compression plates. The reaction forces are recorded. Due to the small forces exerted by each shape memory hub, force-displacement measurements of six specimens are obtained under the same condition. The mean is plotted (Fig.~\ref{fig:4}a). A load cell with a maximum load of \SI{250}{\newton} is used, the measurement error is $\pm 0.005\%$ ($\pm$\SI{0.00125}{\newton}). The same geometry is simulated using finite element analysis. The shape memory hub is modeled using a linear viscoelastic constitutive model~\cite{Chen2017a} and deformed with a prescribed displacement $d=[0-\SI{10}{\milli\metre}]$. The shape memory hub exhibits a non-linear force displacement behavior (Fig.~\ref{fig:4}a) as it bends. Numerical and experimental results are comparable, despite the noise introduced due to the low force output.

The radial pressure is calculated with trigonometry (Fig.~\ref{fig:4}b). As $F_\mathrm{H}$ is horizontal at all stages of loading, the force in the radial direction depends only on the angle $\alpha$. We empirically measure $\alpha$ from the expanded to the collapsed configuration and plot against the radial pressure (Eq.~\ref{eq:pressure}), where $n$ is the number of polygon edge and $R_\mathrm{exp}$ is the outer radius. %This pressure overcomes friction at the pins, and between the system and the ground.

\begin{equation}\label{eq:pressure}
p=F_\mathrm{H}\frac{n}{2\pi R_\mathrm{exp}}\frac{\tan(\alpha)}{\cos(\frac{\pi}{n})}.
\end{equation}

\subsection{Elastic ``flasher'' origami}

Origami structures, like the Miura fold~\cite{Wei2013}, have been studied extensively in the context of deployable systems~\cite{Miura1985}. In the case of rigid origami, the path of folding and unfolding can be predicted with precision~\cite{Tachi2009}. The ability to predict the folding and unfolding paths allows the design of origami mechanisms that can be actuated using motors or thrusters. Reconfigurable systems can benefit from the large volume change achievable by some origami patterns. In existing deployable designs, origami surfaces typically serve as the passive components in a motorized unfolding mechanism~\cite{Mori2010}. Indeed, while the concept of using a Hoberman like scissor mechanism to unfold solar panels has already been proposed~\cite{Hoberman2004}, its actuation or the space to store rigid panels remained an unresolved problem. In additional to origami's folding kinematics, the mechanical behavior of folded origami structures is becoming a focus of study~\cite{Schenk2013}. With the introduction of 4D printing, fold lines within an origami pattern could be programmed to self-fold~\cite{Ge2014a}. This reconfigurability has recently been exploited in the design of acoustic waveguides~\cite{Babaee2016}, prismatic reconfigurable materials~\cite{Overvelde2017} and tunable thermal expansion meta-material~\cite{Boatti2017}.

\begin{figure}[!ht]
	\includegraphics[width=0.45\textwidth]{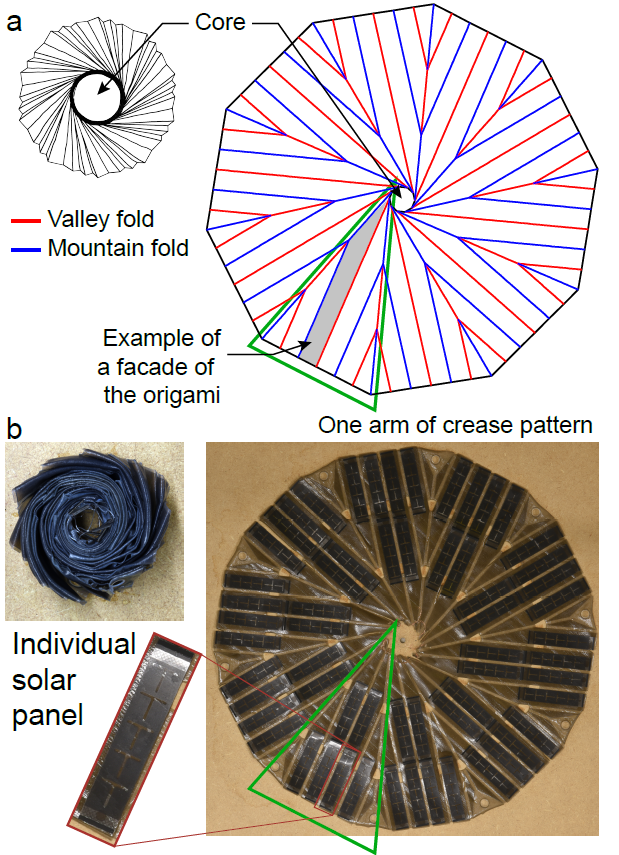}
	\caption{Elastic ``flasher'' origami in both collapsed and expanded configurations. a) The ``flasher'' origami with crease pattern indicated. b) Fabricated specimen with placement of flexible solar panels indicated. The crease pattern is set such that a maximum number of solar panels of a given dimension can fit.}
	\label{fig:5}
\end{figure}

To create a foldable panel with the smallest possible collapsed footprint, we adopt an origami pattern called the ``flasher''(Fig.~\ref{fig:5}a)~\cite{Shafer2001}. The ``flasher'' origami starts as a flat sheet, and can then be folded into a cylindrical element, by allowing each section (arm) to fold around a center core. During the experiments, a mock up of the solar panels are printed on acetate which has comparable stiffness~\cite{Czanderna1996} and same dimension~\cite{FlexSolarCells} of commercially available flexible solar panels (Fig.\ref{fig:5}b).

\begin{figure}[!hb]
	\includegraphics[width=0.45\textwidth]{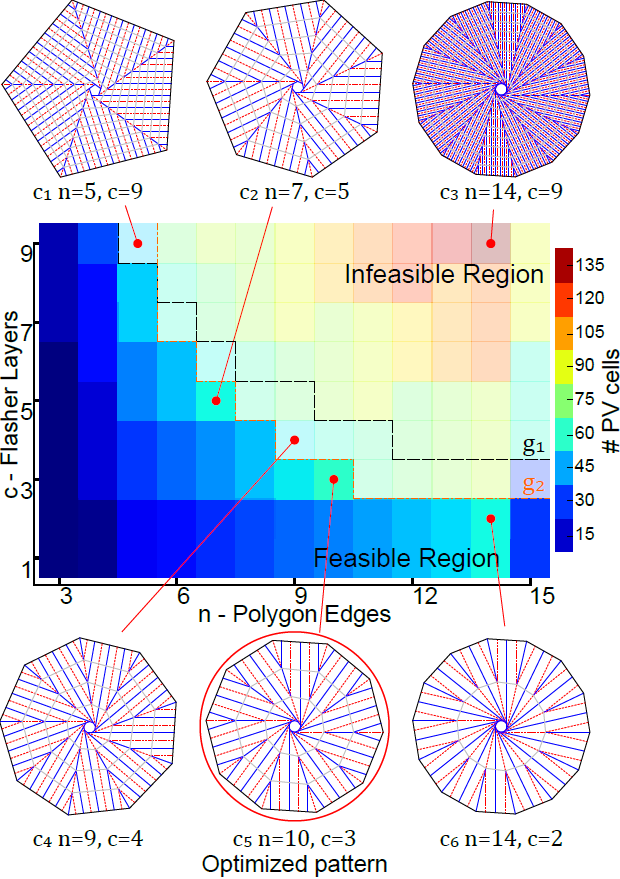}
	\caption{Maximizing the number of flexible solar panels by changing the ``flasher'' origami crease pattern. Two variables, $n$ - number of edges in the pattern's outer polygon, $c$ - number of radial layers in the pattern, are optimized to fit the largest number of solar panels of a given dimension. The contour plot shows the number of solar panels that can fit for every pair of $(n,c)$.}
	\label{fig:6}
\end{figure}

\begin{figure*}[!ht]
	\includegraphics[width=\textwidth]{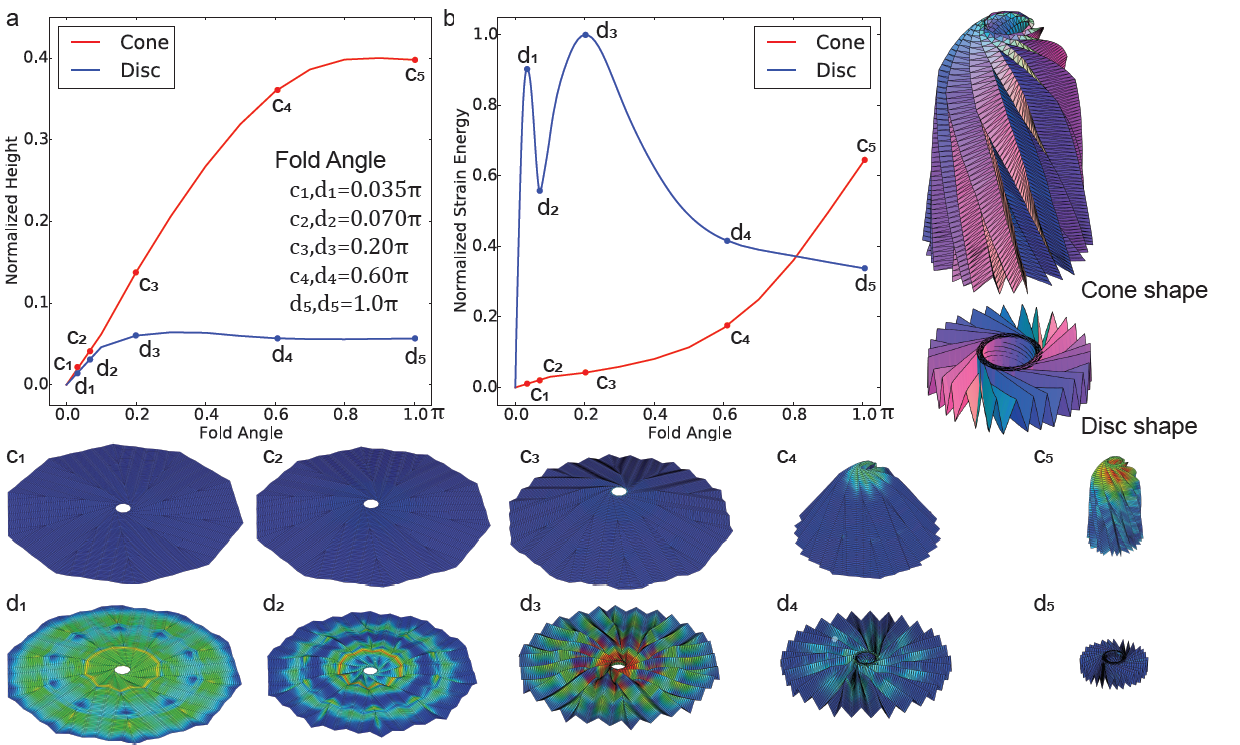}
	\caption{Bifurcation during folding of the ``flasher'' origami as explained by the elastic energy within the fold lines. Two distinct folded shapes are observed during folding both in experiment and in simulation. a) Plot of total height of the origami as a function of the fold angle (between $0$ and $\pi$). b) Plot of the total elastic energy as a function of the fold angle. c) Simulated folded shapes (i.e., $\alpha=\pi$) resulting in either a cone, or a disc.}
	\label{fig:7}
\end{figure*}

Much work has been done on the deployability of rigid foldable versions of the ``flasher'' origami~\cite{Sigel2014, Zirbel2013}. Our flexible crease pattern can be categorized with two variables~\cite{Zirbel2013}. The first variable is the rotational order of the ``flasher'' $n$ which is also the number of sides of the outer polygon. The second variable is the number of circumferential layers in the crease pattern, $c$ . We do not physically discretize the facades in the radial direction, but we allow them to fold flexibly.

As we intend the origami to serve as a hosting platform for solar panels, we optimize the crease pattern to fit the maximum number of solar panels of a given size. While flexible solar panels can bend to a certain curvature, if they are placed across folding lines, they would increase the folding resistance and potentially break during folding. Therefore, they are only placed on the facades in between crease lines. In the generation of the fold pattern, we neglect the thickness of the facades, and allow them to be straight, to better accommodate rectangular solar panels. The strain during folding is assumed to be absorbed by the elasticity of the selected materials. The optimization problem used for design is stated in Eq.~\ref{eq:1}, where we maximize the number of solar panels that can fit within a given pattern. Constraints to this problem are as follows, 1) the outer dimension of the unfolded and folded origami must fit within the expanded and collapsed Hoberman ring respectively, and 2) Each facade of the crease pattern must be at least as wide as a given solar panel plus the width of the fold lines. Preference is given to patterns that can construct periodic linkages to the Hoberman ring.

\begin{equation}
 \begin{aligned}
& \underset{c,n}{\text{minimize}} && -\max{\left[p(c,n),0\right]}\\
& \text{where}\\
& && p(c,n)=\mathrm{floor}\left[\sum_{i=1}^{c}\left({r_\mathrm{u,o}-\frac{w_\mathrm{PV}+h_\mathrm{f}(i-1)}{\tan(\pi / n)}}\right)\right] \\
& \text{subject to}\\
& && r_\mathrm{f,o} - r_\mathrm{max}\leq 0\\
& && w_\mathrm{PV}-h_\mathrm{f}\leq 0\\
\label{eq:1}
\end{aligned}
\end{equation}

Where $r_\mathrm{u,o}$ and $r_\mathrm{f,o}$ are the outer polygon radii of the unfolded and folded configuration respectively, $w_\mathrm{PV}$ is the smaller dimension of a rectangular solar panel, $h_\mathrm{f}$ is the height of the origami in the folded configuration, $r_\mathrm{max}$ is the inner void space of the Hoberman ring (as discussed in the previous section, and detailed in Appendix I).
 
The feasible region of the objective landscape (Fig.~\ref{fig:6}) show optimized solutions of 60 solar panels at $(n,c)=(7,5), (10,3), (14,1), (14,2)$. As expected, as the number of rotational order $n$ increase, more solar panels can be fitted. The constraints however dictate that beyond a certain $n$ or $c$, the folded dimension $r_\mathrm{f,o}$ exceeds that of the inner dimension of the Hoberman ring $r_{\mathrm{max}}$. The adopted solution is a decagon ($n=10$) with three layers ($c=3$).We adopt this design because $n=10$ is a fraction of the number of scissor mechanisms in the Hoberman ring. In this way, periodic connections can be made between the two. The derivation of the folded dimension of the origami is detailed in Appendix II.

The design is fabricated with two layers: The 3D printed base layer and the solar panel layer with thicknesses $\SI{0.15}{\milli\metre}$ and $\SI{0.24}{\milli\metre}$, respectively. The thickness of the facades provide structural rigidity during the initial folding process, but is flexible enough to be wound around the center core. The gaps between the facades form the lines of fold. The gaps width is greater than $2t$ where t is the thickness of the facades. Due to the inherent elasticity of the material in the rubbery state, both mountain and valley crease lines are assigned the same width. Voids are placed at the intersection of multiple crease lines to reduce stress concentration.

\begin{figure*}[!ht]
	\includegraphics[width=\textwidth]{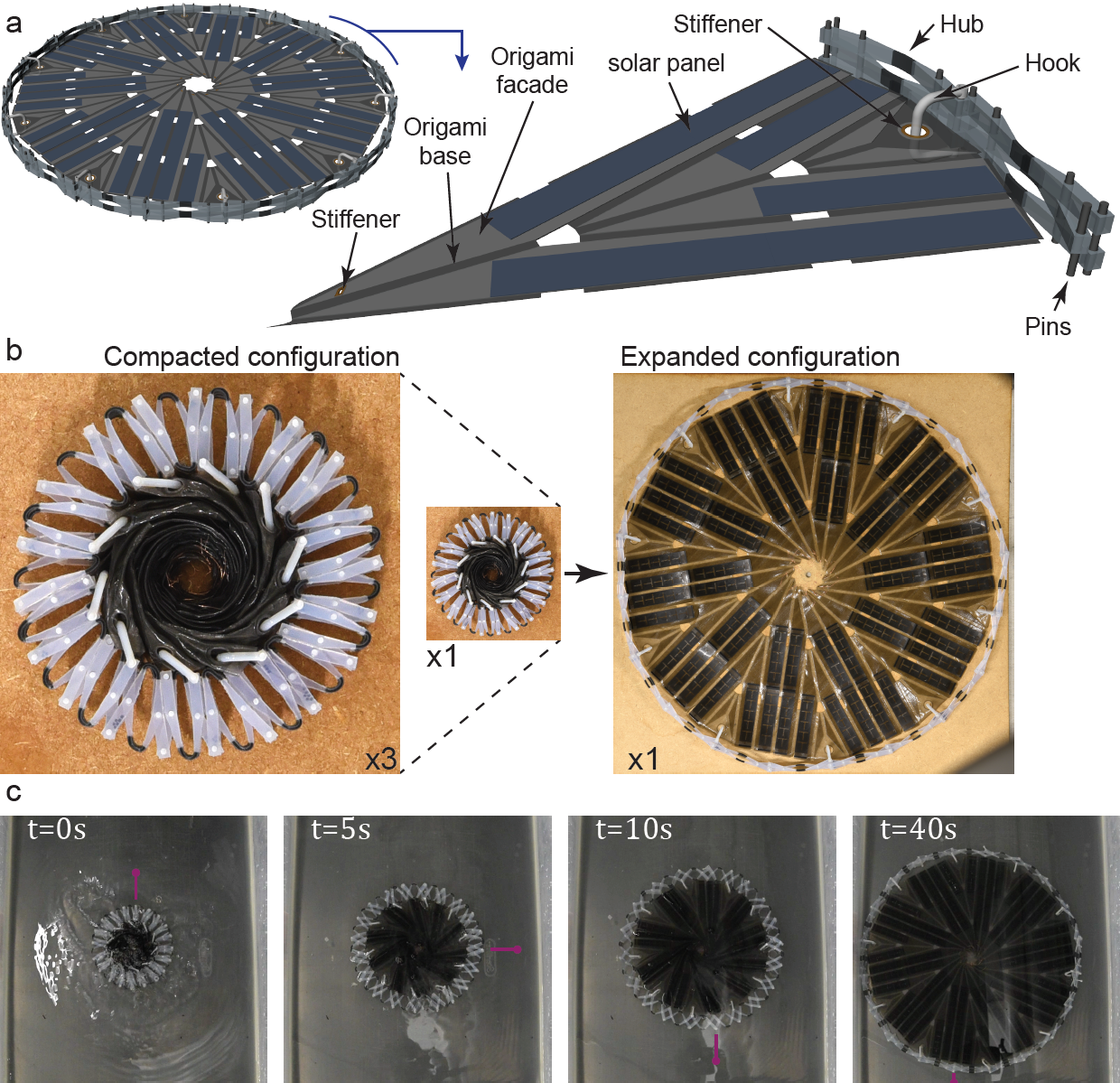}
	\caption{Deployable solar panel array. a) Rendering of the design assembly with the different components labeled. b) Photos showing the collapsed and the expanded configuration of the solar panel array. c) Video frames of the deployment process in water above $T_\mathrm{g}$.}
	\label{fig:8}
\end{figure*}

For the programming of the origami and consequently of the whole assembly, we apply a rotation at the center core while keeping the mountain and valley fold lines of the origami in the correct half plane(i.e. $z>0$ and $z<0$)~\cite{Lanford1961}. To apply this rotation, flexible wires are connected between the inner most vertices of the origami and a center rotational core (See Fig.~\ref{fig:SI_4}). It is noted that during folding, the center may ``pop'' out of plane, forcing the whole origami to irreversibly assume a cone-like shape. To better understand this phenomenon, the folding process is simulated by assuming the crease pattern is an edge-node network. The coordinates of the nodes are solved such that there is minimal axial strain along the bars, and that the dihedral fold angle between all crease lines equal to a prescribed value $\theta$~\cite{Schenk2011}. By incrementing $\theta$, a pseudo-dynamic folding simulation is achieved~\cite{Ghassaei2017}. As this method is used for rigid origami, the facades of the ``flasher'' in the radial direction are finely segmented to mimic a flexible folding behavior. %The error is then the elastic energy originated from strain within the bars, as this strain reflects stretching or compression of the panels.
 A bifurcation behavior is observed, i.e. two different heights can be achieved from the same fold angle (Fig.~\ref{fig:7}a, $c_{1-5}$ and $d_{1-5}$). The cone shape is triggered if the change in the folding angle is small between two pseudo time steps, and the disc shape is triggered if the change is large. We see that the strain energy or the error is much larger for smaller fold angles for the disc than it is for the cone. This correlation inverses at a larger fold angle (Fig.~\ref{fig:7}b). This explains the center of the origami ``popping-up'' during folding, and is avoided in experiments by manually holding the center of the ``flasher'' down.% and folding in shallow water to minimize buoyancy.

\subsection{Autonomous self-deploying solar panel array}
While the proposed design can be fabricated using a 3D printer monolithically, it is made as an assembly for demonstration purposes. Both the Hoberman ring and the elastic origami are fabricated in the expanded configuration and assembled using rotational hooks(Fig.~\ref{fig:8}a). While both components have been programmed individually for demonstration (Fig.~\ref{fig:2}c and Fig.~\ref{fig:5}b), the assembly is programmed in one step in a heated environment ($T\ge T_{\mathrm{g}}$). The folding is done by a single person, in contrast to the common practice of large origami patterns that usually requires as many people as the number of arms. The collapse starts by rotating the core which the facades are connected to. As the core rotates, the whole geometry is pulled in and wrapped. Then a cylindrical mold that would give the desired collapsing ratio is used to demonstrate that the collapsed design can fit within a given space. While confined in this mold, the collapsed design is cooled to the stowing temperature (Fig.~\ref{fig:8}b). The mold is then removed.

The design must be deployed in an environment heated past $T_\mathrm{g}$. In our experiments, it is deployed under water, to simulate a reduced gravity environment. First a rapid rotational motion is observed, where the deployment is largely initiated by the origami itself. At a certain stage, the rotation stops and the remaining folds are flattened by the Hoberman ring (Fig.~\ref{fig:8}c). This deployment behavior mimics a proposed two-stage mechanized deployment of a solar sail,\cite{Mori2010} yet requires no actuators or power supply.

\section{Conclusion}
We exploit the large area expansion capability of the Hoberman ring to autonomously deploy an array of flexible solar panels mounted on an elastic ``flasher'' origami substrate. The design of the system is created by analyzing the Hoberman ring and the ``flasher'' origami separately. First we propose to replace the rotational hubs of the Hoberman ring with shape memory polymers to enable temperature-triggered deployment. We then parametrize and optimize the expansion ratio of the Hoberman ring assuming rigid behavior. Lastly, we analyze the mechanics of collapse and expansion of a single shape memory hub.

The analysis of the ``flasher'' origami begins by parameterizing and optimizing the crease pattern to accommodate the largest number of solar panels. Second, a simplified strain-energy model is proposed to explain the seemingly bifurcating behavior exhibited by the structure while folding.

As our proposed design is under the highest mechanical stress during the collapse process, we guarantee that during deployment no material failure can occur. We demonstrate the controlled deployment of a mechanical system without external actuators, sensors or power supply. Rather, it is achieved by embedding and distributing these functionalities within the material itself. 

\section*{Acknowledgments}
C.D. acknowledges support from the Army Research Office Grant W911NF-17-1-0147 and the  Caltech/JPL President's and Director's Fund Program.

\appendix

\begin{figure*}[ht]
 \includegraphics[width=\textwidth]{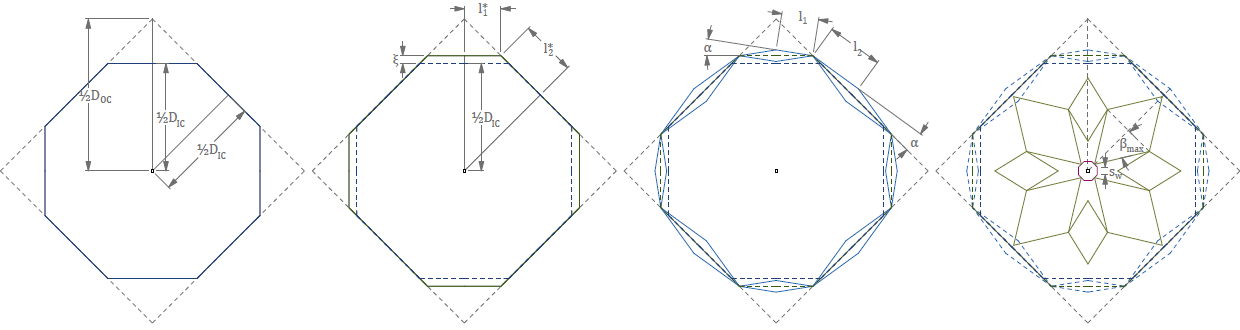}
\caption{Construction of a Hoberman ring. Variables used in the derivation are shown.}
\label{fig:SI_1}
\end{figure*}

\section{Derivation of area expansion ratio of the Hoberman ring}
The Hoberman sphere is created using a regular polyhedron as a base, similarly, the ring can be created using a regular polygon. We derive the analytical relationship between the change in area from the expanded to the collapsed configuration of the Hoberman ring in relation to the number of edges of the initial polygon. Further, we introduce three variables to account for fabrication constraints (Fig.~\ref{fig:SI_1}). Given the circumscribing diameter of the polygon, $d_\mathrm{OC}$, as dictated by the maximum fabrication dimension, we calculate the edge length $s$, the inscribing diameter $d_\mathrm{IC}$, the external angle $\theta_\mathrm{EA}$ and the internal angle $\theta_\mathrm{IA}$.
\begin{equation}
s=d_\mathrm{OC}\sin(\frac{\pi}{n})~~;~~d_\mathrm{IC}=s\cot(\frac{\pi}{n})
\end{equation}

\begin{equation}
\theta_\mathrm{EA}=\frac{2\pi}{n}~~;~~\theta_\mathrm{IA}=(n-2)\frac{\pi}{n}
\end{equation}

We introduce an isogonal polygon with $2n$ edges for the construction of the double scissor mechanism. The edges of this polygon are alternatively grouped into two sets $e_1$ and $e_2$ with lengths $l_1^*$ and $l_2^*$ respectively. The perpendicular distance from the origin to edge set $e_1$ is defined by varying the inscribing diameter $d_\mathrm{IC}$ with $\xi$ to create the isogonal polygon. Edge set $e_2$ rests on the original regular polygon.
\begin{equation}
a=\frac{1}{2}d_\mathrm{IC}+\xi
\end{equation}

\begin{figure*}[ht]
 \includegraphics[width=0.45\textwidth]{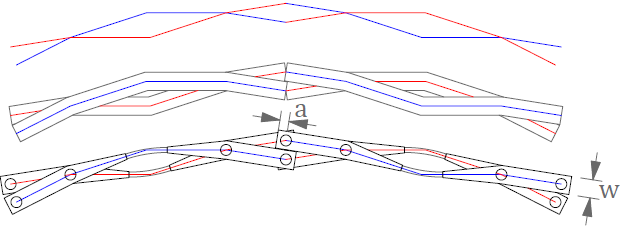}
\caption{Rationale for two different lengths in the scissor segments, the pin ends are longer than the hub ends.}
\label{fig:SI_2}
\end{figure*}

This is done such that the nominal length of scissors with pin ends are shorter than the one with hub ends. The pin ended scissors require an added length as a hole must be made at the end of the link,(Fig.~\ref{fig:SI_2}), where as the hub is a flexible polymer which deforms. By adding $\xi$, we can tailor the system such that the physical embodiment would be closer to equal length. Length of edge set $e_1$ is then,
\begin{equation}
l_1^*=(\frac{1}{2}d_\mathrm{OC}-a)\tan(\frac{\theta_\mathrm{IA}}{2})
\end{equation}

To calculate the length of the edge set $e_2$ we first calculate the two angles, $\theta_1$ and $\theta_2$, formed by connecting the ends of each set of edges to the center,
\begin{equation}
\theta_1=\tan^{-1}\left(\frac{l_1^*}{a}\right),
\end{equation}

\begin{equation}
\theta_2=\frac{1}{2}\theta_\mathrm{EA}-\theta_1.
\end{equation}

Then we find the length of edge set $e_2$,
\begin{equation}
l_2^*=\tan(\theta_2)\frac{d_\mathrm{IC}}{2}.
\end{equation}

If the scissor mechanism has an initial angle $\alpha$,(Fig.~\ref{fig:2}b) the lengths of the half scissor become $l_1$ and $l_2$,
\begin{equation}
l_1=l_1^*\frac{1}{\cos(\alpha)}~~;~~l_2=l_2^*\frac{1}{\cos(\alpha)}.
\end{equation}

Without loss of generality, we assume $\xi\ge0$ since a negative $\xi$ of the same magnitude would only mirror the structure. With this assumption, we see that $l_1<l_2$. We define $l=l_2$ and $\theta=\theta_2$. Now we assign physical dimensions to the collapsed ring. Assuming that the width of each scissor, $w$, equals the edge length of a fictitious inner polygon of edge $2n$, and an inscribing diameter $d_\mathrm{IC, f}$, then we can calculate an angle $\beta_\mathrm{max}$ as the angle between two scissors in the collapsed configuration:
\begin{equation}
d_\mathrm{IC, f}=w\cot(\frac{\pi}{2n}).
\end{equation}

The maximum angle of rotation between the scissors is therefore, 
\begin{equation}
\beta_\mathrm{max}=\frac{\pi}{2}-\theta-\arcsin{\frac{d_\mathrm{IC,f}\sin(\theta)}{2l}}.
\end{equation}

We define the ratio of area expansion as:
\begin{equation}
\Delta A=\frac{\left[d_\mathrm{IC}+2\sin(\alpha)l\right]^2}{\left[4l\sin(\beta_\mathrm{max})+d_\mathrm{IC, f}\right]^2}.
\end{equation}

In figure 1 of the manuscript, we plot $\Delta A$ vs. $n$ for different values of $w$.

\section{Optimization landscape and constraint boundaries of the ``flasher'' origami}
In this work, we adopt the design of a ``flasher'' origami as the basis to attach the solar panels. We relate the shape and folds of a flash origami pattern to the number of solar panels that can fit on such a pattern. With this, we are able to generate a crease pattern that can accommodate the given number of solar panels of a certain size. We also use this derivation to relate the radius of the unfolded pattern to the folded one.

\begin{figure*}[ht]
 \includegraphics[width=\textwidth]{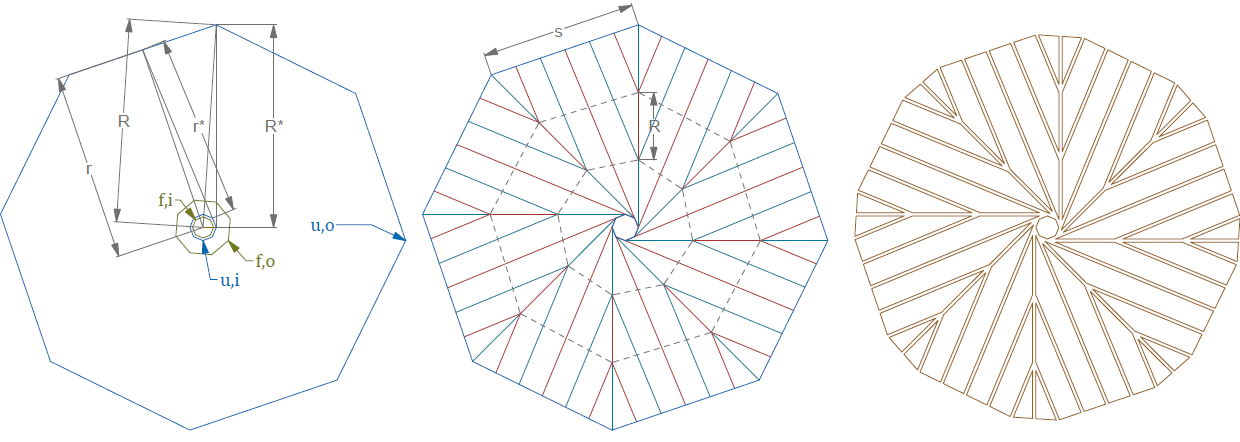}
\caption{Construction of a crease pattern of a ``flasher'' origami. Variables used in the derivation are shown.}
\label{fig:SI_3}
\end{figure*}

Given two concentric polygons of $n$ sides, we rotate the outer polygon such that the edges of the inner polygon, if extended outward, bisects the edges of the outer polygon. This defines the outer and inner limits of the unfolded pattern. Two other polygons of $n$ sides are used to define the outer and the inner limits of the folded pattern. The circumscribing radius, inscribing radius, and edge length are denoted as $R$, $r$, and $s$ respectively. 

The superscript $^*$ refers to two outer polygons offset by the asymmetry of the origami pattern (Fig.~\ref{fig:SI_3}). The relationship between the offset quantities is,
\begin{equation}
R_{u,o}={({{R_{u,o}}^*}^2+R_{u,i}^2)}^{\frac{1}{2}}
\end{equation}

These four polygons are denoted with subscripts $_{u,o}$, $_{f,o}$, $_{u,i}$, $_{f,i}$ referring to unfolded outer, folded outer, unfolded inner, and folded inner respectively. The known quantities are $R_{f,i}$, ${R_{u,o}}^*$, $R_{u,i}$.

 We define the radius of the folded geometry, $R_{f,o}$, as $R_{u,i}$ plus the number of wrapped layers, $p$. The variables $c$ and $t$ refers to the number of levels in the ``flasher'' pattern, and the thickness of origami sheet respectively.
\begin{equation}
R_{f,o}=R_{f,i}+p(2ct)
\end{equation}

The number of edges each arm of the outer polygon can wrap around the inner polygon is calculated as the ratio between the inscribed radius of the unfolded geometry and the edge length of the folded geometry, $S_{f,o}$. Note that this edge length increases as more layers are wrapped and the radius increases.
\begin{equation}
p=\frac{{r_{u,o}}^*}{s_{f,o}}
\end{equation}

The inscribed radius is $r_{u,o}=R_{u,o}\cos{\frac{\pi}{n}}$ and the edge length is $s_{u,o}=2R_{u,o}\sin{\frac{\pi}{n}}$. Similarly, we calculate the edge lengths $s_{f,i}$, $s_{u,i}$, and the inscribed radius of the inner polygon $r_{u,i}$. we calculate the offset inscribed radius,
\begin{equation}
{r_{u,o}}^*={({r_{u,o}}^2-{r_{u,i}}^2)}^{\frac{1}{2}}-\frac{s_{u,i}}{2}
\end{equation}

and the edge length of the folded polygon follows,
\begin{equation}
s_{f,o}=2R_{f,o}\sin(\frac{\pi}{n}).
\end{equation}

We can solve for $R_{f,o}$, which forms the $g_2$ constraint in Figure~\ref{fig:2} of the manuscript.

\begin{widetext}
\begin{equation}
\begin{aligned}
R_{f,o}=& \left\{ \sqrt {\sin \left( {\frac {\pi}{n}} \right) \left( -4ct\sin
 \left( {\frac {\pi}{n}} \right) { R_{u,i}}+4ct\sqrt { \left( 
\cos \left( {\frac {\pi}{n}} \right) \right) ^{2}{{ R_{u,o}^*}}^{2}}+\sin \left( {\frac {\pi}{n}} \right) {{ R_{f,i}}}^{2} \right) }{
 R_{f,i}} \right. \\
& \left . +2ct\sqrt { \left( \cos \left( {\frac {\pi}{n}} \right) 
 \right) ^{2}{{ R_{u,o}^*}}^{2}}+ \left( -2ct{ R_{u,i}}+{{ 
R_{f,i}}}^{2} \right) \sin \left( {\frac {\pi}{n}} \right) \right\} \cdot \\
& \left\{ { R_{f,i}}\sin \left( {\frac {\pi}{n}} \right) +\sqrt {\sin
 \left( {\frac {\pi}{n}} \right) \left( -4ct\sin \left( {\frac {\pi}{
n}} \right) { R_{u,i}}+4ct\sqrt { \left( \cos \left( {\frac {\pi
}{n}} \right) \right) ^{2}{{ R_{u,o}^*}}^{2}}+\sin \left( {\frac {
\pi}{n}} \right) {{ R_{f,i}}}^{2} \right) } \right\} ^{-1}
 \end{aligned}
\end{equation}
\end{widetext}
The first constraint requires the height of the folded origami to be greater than the smaller dimension of the solar panels. The height is simply the edge length of the unfolded polygon divided by twice number of layers in the ``flasher'':
\begin{equation}
h_f={\frac {\sqrt {{{R_{u,i}}}^{2}+{{R_{u,o}^*}}^{2}}}{c}\sin \left( {\frac {\pi}{n}} \right) }.
\end{equation}

\section{Collapsing procedure}
Thermomechanically, the procedure to collapse the system from the deployed configuration to the stowed configuration is referred to as ``programming''. It is done at a temperature higher than the glass transition temperature of the shape memory polymers within the system.

\begin{figure*}[hb]
 \includegraphics[width=\textwidth]{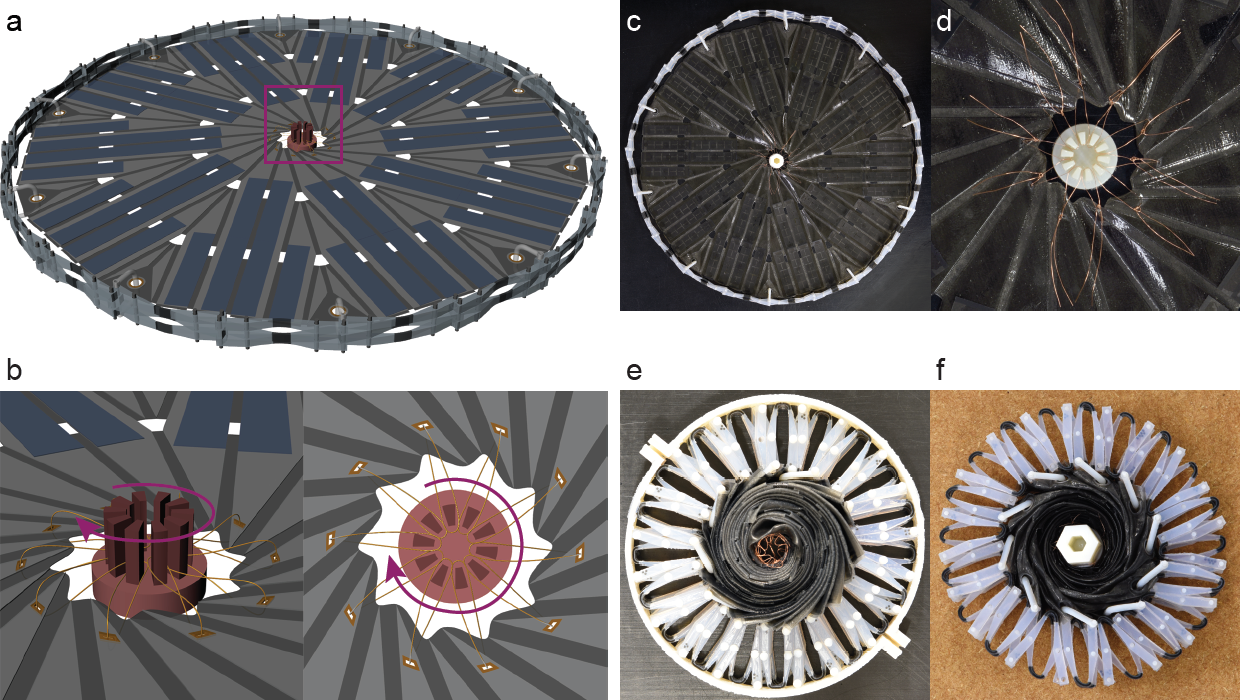}
\caption{``Programming'' of the solar panel array. a,c) The procedure starts with the deployed configuration. b,d) A center rotational core is installed and connected with the origami via flexible wires. e) In a heated environment, the system is rotated and folded until it fits within a cylindrical mold. f) After cooling within the mold, it is released and stable on its own.}
\label{fig:SI_4}
\end{figure*}

Components of the system (i.e. the ``flasher'' origami, the scissor mechanisms, and the connections between the two) are fabricated in the deployed configuration and assembled along with the solar panels (Fig.~\ref{fig:SI_4}a,c). A center core is installed and connected to the ``flasher'' origami through a number of flexible wires (Fig.~\ref{fig:SI_4}b,d). 

The assembled system is submerged in heated water ($T\ge T_\mathrm{g}$), and the center core is rotated as the crease lines are folded. This continues until the collapsed system fits within a prescribed cylinder (Fig.~\ref{fig:SI_4}e). Once in the cylinder and cooled to below the glass transition temperature, the cylinder is released (Fig.~\ref{fig:SI_4}f).

\end{document}